\documentclass[12pt]{article}
\usepackage{amsmath, amssymb, graphicx}
\usepackage[affil-it]{authblk}
\usepackage{amsthm}
\usepackage{booktabs}
\usepackage{adjustbox}
\newtheorem{theorem}{Theorem}
\newtheorem{lemma}{Lemma}
\usepackage[margin=0.8in]{geometry}
\usepackage{url}
\usepackage{hyperref}
\title{Bayesian Inference for Left-Truncated Log-Logistic Distributions for Time-to-event Data Analysis}

\author[1]{Fahad Mostafa}
\author[2]{Md Rejuan Haque}
\author[3]{Md Mostafijur Rahman}
\author[4]{Farzana Nasrin}
\affil[1]{NCIAS, Arizona State University, Phoenix, AZ 85306}
\affil[2]{Center for Biostatistics, Department of Biomedical Informatics, The Ohio State University, Columbus, OH 43210}
\affil[3]{Department of Electrical and Computer Engineering, The University of Texas at Austin, Austin, TX 78712}

\affil[4]{Department of Mathematics, The University of Hawaii at Mānoa, HI 96822}

\date{}

\begin{document}

\maketitle

\begin{abstract}
Parameter estimation is a foundational step in statistical modeling, enabling us to extract knowledge from data and apply it effectively. Bayesian estimation of parameters incorporates prior beliefs with observed data to infer distribution parameters probabilistically and robustly.
Moreover, it provides full posterior distributions, allowing uncertainty quantification and regularization, especially useful in small or truncated samples. Utilizing the left-truncated log-logistic (LTLL) distribution is particularly well-suited for modeling time-to-event data where observations are subject to a known lower bound such as precipitation data and cancer survival times. In this paper, we propose a Bayesian approach for estimating the parameters of the LTLL distribution with a fixed truncation point \( x_L > 0 \). Given a random variable \( X \sim LL(\alpha, \beta; x_L) \), where \( \alpha > 0 \) is the scale parameter and \( \beta > 0 \) is the shape parameter, the likelihood function is derived based on a truncated sample \( X_1, X_2, \dots, X_N \) with \( X_i > x_L \). We assume independent prior distributions for the parameters, and the posterior inference is conducted via Markov Chain Monte Carlo sampling, specifically using the Metropolis-Hastings algorithm to obtain posterior estimates \( \hat{\alpha} \) and \( \hat{\beta} \). Through simulation studies and real-world applications, we demonstrate that Bayesian estimation provides more stable and reliable parameter estimates, particularly when the likelihood surface is irregular due to left truncation. The results highlight the advantages of Bayesian inference outperform the estimation of parameter uncertainty in truncated distributions for time to event data analysis.

\end{abstract}

\textbf{keywords} Bayesian estimation, matching prior, left-truncated Log-logistic distribution, credible interval, time-to-event datasets. 
\section{Introduction}

Time-to-event data require specialized techniques to handle censoring and time-dependence \cite{thompson2010statistical}. Their behavior is nuanced, shaped by the distribution of event times and censoring patterns such as statistical modeling with LTLL distribution. However, parameter estimation is a complex task that requires precise modeling, robust algorithms, and careful handling of data uncertainties to ensure accurate, reliable inference in diverse scientific and engineering applications. Among many techniques, 
Bayesian methods work well even when the sample size is small, which is often a challenge in real-world truncated data scenarios \cite{bon2023being}. Prior distributions help compensate for limited data. Bayesian estimation provides a full posterior distribution for the parameters rather than single point estimates. This means one gains deeper insight into the uncertainty and credibility of parameter estimates, which is critical for truncated data \cite{gelman2013bayesian}. Moreover, Bayesian parameter estimation is particularly effective when applying the LTLL distribution in survival analysis or reliability engineering. For example, when modeling time-to-failure data in systems where early failures are excluded due to truncation, Bayesian methods can provide a robust framework for estimating the scale, shape, and truncation parameters \cite{umlauf2018bamlss}. Furthermore, by incorporating prior knowledge, such as historical reliability data or expert judgment about the system's performance, Bayesian estimation improves accuracy, especially in scenarios with limited or incomplete datasets. Since the log-logistic distribution is widely used in survival analysis, economics, and hydrology. The log-logistic distribution, also referred to as the Fisk distribution, has been widely utilized across various fields due to its mathematical tractability and suitability for modeling skewed data. Originally introduced in the econometric literature by Fisk in 1961 \cite{fisk1961estimation, fisk1961graduation}, this distribution gained recognition for its superior ability to represent income distributions compared to the Pareto model. Beyond economics, the log-logistic distribution has been extensively applied in hydrology and environmental sciences. For instance, Shoukri et al. \cite{shoukri1988sampling} demonstrated its effectiveness in modeling Canadian precipitation data, while Ahmad et al. \cite{ahmad1988log} utilized it for estimating flood frequencies in Scotland. More recently, the distribution has been applied in reliability analysis and survival studies due to its flexibility in capturing hazard rate variations by Murthy et al. \cite{Murthy2004}, and Wang \& Kececioglu \cite{Kececioglu2002}. Mathematically, the log-logistic distribution is closely related to the logistic distribution via a logarithmic transformation, making it a convenient choice for analytical derivations. Its widespread adoption is largely attributed to its closed-form expressions for both the cumulative distribution function, and probability density function, which simplify parameter estimation and inference by Reath et al. \cite{Reath2018}, He et al.\cite{He2020}. In addition, its ability to accommodate heavy-tailed distributions makes it a valuable tool in financial risk modeling, biomedical research, and reliability engineering by Johnson et al. \cite{Johnson1995}, and Balakrishnan \& Kundu \cite{Balakrishnan1991}. When log-logistic distribution is left-truncated, the estimation of its parameters becomes more challenging. Bayesian estimation provides an alternative that incorporates prior knowledge and regularization, leading to more stable estimates. Bayesian estimation of the parameters provides an alternative that incorporates prior knowledge and regularization, leading to more stable estimates \cite{singh2013bayesian, swaminathan1985bayesian}. Recently, Jaakkola et al. explored Bayesian parameter estimation using variational techniques. Their work shows that the dual formulation of the regression problem results in a latent variable density model, and the corresponding variational approach yields EM updates that can be solved exactly \cite{jaakkola2000bayesian}.  Moreover, Bayesian parameter estimation has been widely applied across various distribution families, providing a flexible framework for statistical inference under uncertainty. For example, recent work has focused on estimating the scale parameter in the Erlang distribution using informative priors, yielding more accurate results compared to classical approaches \cite{Abdelfattah2023}. Similarly, the three-parameter Weibull distribution has been analyzed using Bayesian techniques for improved reliability analysis \cite{Zhou2023}. Bayesian methods have also been utilized in estimating parameters of inverse exponentiated distributions, with estimators derived under different loss functions, offering robustness and interpretability \cite{Tahmasbi2023}. Furthermore, Bayesian inference has been applied to inverse Gaussian distributions, particularly in modeling life data and estimating reliability measures \cite{Zhang2024}. Finally, complex models such as the three-component mixture geometric distribution have been addressed via Bayesian estimation, demonstrating the approach's capability to handle mixture distributions effectively \cite{Paul2024}.\\

\noindent This study introduces a Bayesian framework for estimating parameters of the left-truncated log-logistic distribution, offering improved robustness, and addressing uncertainties. Our findings demonstrate that Bayesian estimation provides more stable and accurate parameter estimates. Through theoretical derivations and empirical analyses, we establish the conditions under which Bayesian and  inference as well as illustrate its advantages using real-world datasets. The introduction of error ellipses further quantifies the uncertainty in parameter estimates, highlighting the benefits of Bayesian regularization. These findings provide new insights into parameter estimation for truncated distributions, making Bayesian methods a preferred choice in practical applications involving left-truncated log-logistic distributions. The whole paper is organized as follows. In Section \ref{sec:2}, the Bayesian estimation of the parameters is introduced, and the LTLL distribution is defined. In Section \ref{sec:3}, the Bayesian framework for the LTLL distribution is presented. Theoretical properties of the proposed methodology are analyzed in Section \ref{sec:4}. In Section \ref{sec:5}, the computational algorithm for Bayesian inference is described. Finally, applications to simulated and real data using the proposed method are presented in Section \ref{sec:6} and \ref{sec:7}.

\section{Background}\label{sec:2}

Bayesian estimation provides a probabilistic framework for parameter inference by combining prior beliefs with observed data using Bayes' theorem \cite{gelman2013bayesian}. Mathematically, let \( \theta \in \Theta \subseteq \mathbb{R}^d \) denote the vector of unknown parameters and let \( X = \{x_1, x_2, \dots, x_n\} \) be the observed time-to-event data, assumed to follow a likelihood function \( L(\theta) = p(X \mid \theta) \). In the Bayesian framework, the parameter \( \theta \) is treated as a random variable with a prior distribution \( \pi(\theta) \). Using Bayes' theorem, the posterior distribution of \( \theta \) given the data is expressed as:
\[
\pi(\theta \mid X) = \frac{p(X \mid \theta) \pi(\theta)}{\int_{\Theta} p(X \mid \theta) \pi(\theta) \, d\theta} = \frac{L(\theta) \pi(\theta)}{m(X)},
\]
where \( m(X) = \int_{\Theta} L(\theta) \pi(\theta) \, d\theta \) is the marginal likelihood. Bayesian parameter estimates are obtained by summarizing the posterior; for instance, the posterior mean is

\begin{equation*}
    \hat{\theta}_{\text{Bayes}} = \mathbb{E}[\theta \mid X] = \int_{\Theta} \theta \, \pi(\theta \mid X) \, d\theta,
\end{equation*}
and a \( 100(1 - \gamma)\% \) credible interval \( (l_\gamma, u_\gamma) \) satisfies
\begin{equation*}
    \mathbb{P}(l_\gamma \leq \theta \leq u_\gamma \mid X) = 1 - \gamma.
\end{equation*}
This approach yields not only point estimates but also a full posterior distribution for assessing parameter uncertainty. From above, the point estimates, such as the posterior mean $ \mathbb{E}[\theta \mid X]$, and interval estimates, such as credible intervals, can be derived directly from the posterior, offering a comprehensive view of parameter uncertainty (see in \cite{gelman2013bayesian, box1965bayesian}).

\begin{figure}[h!]
    \centering
    \includegraphics[width=0.9\textwidth]{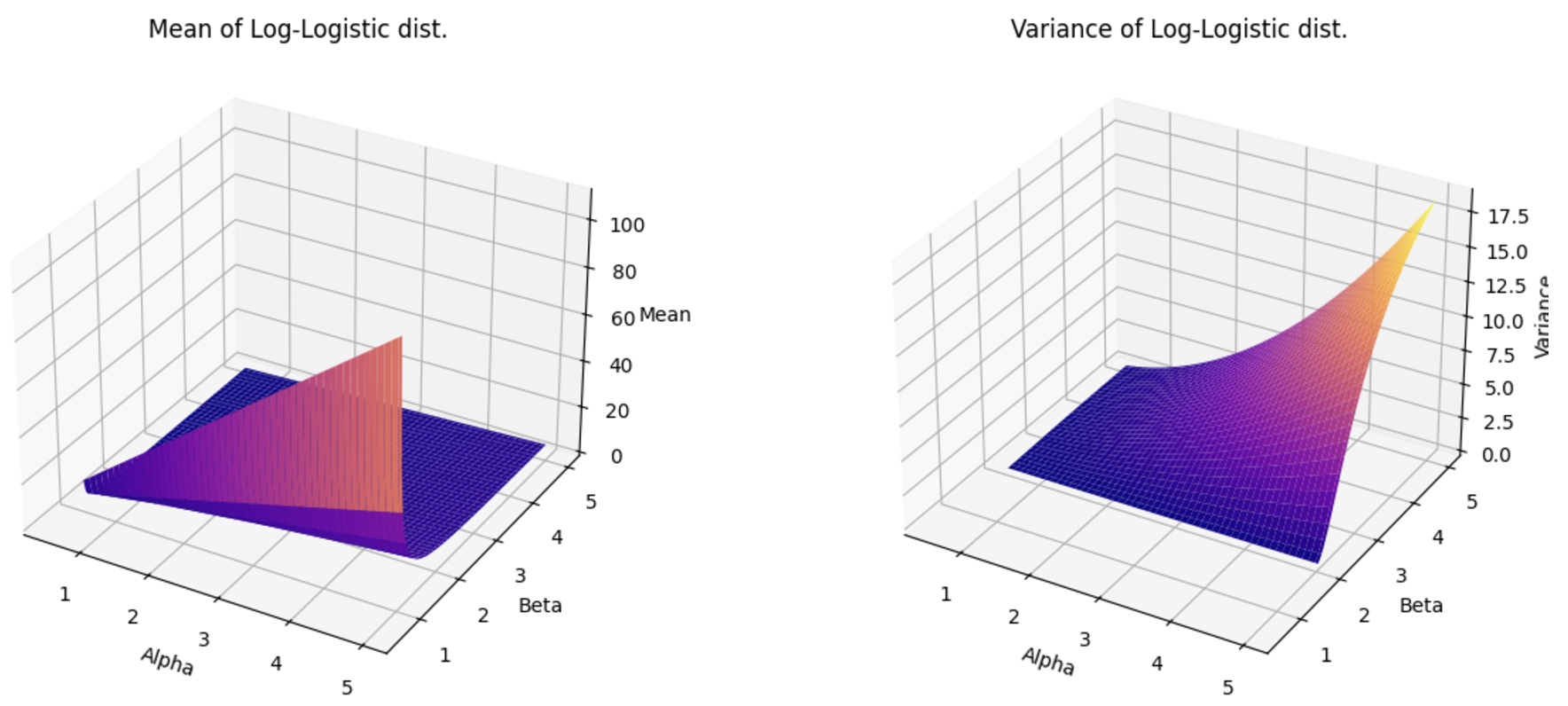}
    \caption{Moment surfaces of the log-logistic distribution showing how the mean, and variance vary with parameters $\alpha$ and $\beta$.}
    \label{fig:ll_moments}
\end{figure}

The log-logistic distribution, denoted as $LL(\alpha, \beta)$, is characterized by its PDF:

\begin{equation}
    f(x | \alpha, \beta) = \frac{\beta}{\alpha} \left( \frac{x}{\alpha} \right)^{\beta - 1} \left[ 1 + \left( \frac{x}{\alpha} \right)^{\beta} \right]^{-2}, \quad x > 0.
\end{equation}

where $\alpha > 0$ is the scale parameter and $\beta > 0$ is the shape parameter (see Figure \ref{fig:ll_moments}). For a left-truncated log-logistic distribution with truncation point $x_L > 0$, the PDF is given by:

\begin{equation*}
    f_{LT}(x | \alpha, \beta; x_L) = \frac{f(x | \alpha, \beta)}{1 - F(x_L | \alpha, \beta)}, \quad x > x_L.
\end{equation*}

where the cumulative distribution function is:
\[
F(x | \alpha, \beta) = \frac{1}{1 + (x/\alpha)^{-\beta}}.
\]

A crucial property of the log-logistic distribution is its ability to model data with heavy tails, making it suitable for applications in survival analysis and economics (see e.g. \cite{gupta1999study, maurya2021new}). Its flexibility in capturing extreme events has made it a popular choice in modeling phenomena such as annual flood maxima \cite{ahmad1988log, coles2001introduction, tasker1998regional}. Traditionally, a threshold parameter is introduced to ensure that observed values exceed a minimum level. In this study, we instead use a fixed left-truncation point $x_L > 0$, leading to the left-truncated log-logistic distribution with probability density function \cite{cohen1991truncated, kreer2024maximum}:

\begin{equation*}
    f_{LT}(x | \alpha, \beta; x_L) = \left( 1 + \left( \frac{x_L}{\alpha} \right)^\beta \right) \frac{\beta}{\alpha} \left( \frac{x}{\alpha} \right)^{\beta-1} \frac{1}{\left[ 1 + \left( \frac{x}{\alpha} \right)^\beta \right]^2},
\end{equation*}

and cumulative distribution function is:

\begin{equation*}
    F_{LT}(x | \alpha, \beta; x_L) = \frac{\left( \frac{x}{\alpha} \right)^\beta - \left( \frac{x_L}{\alpha} \right)^\beta}{1 + \left( \frac{x}{\alpha} \right)^\beta}.
\end{equation*}

If a random variable $X$ follows an LTLL distribution, denoted as $X \sim LL(\alpha, \beta; x_L)$, it can be generated from a uniform random variable $U \sim \text{Uniform}(0,1)$ using:

\begin{equation}
    X = \alpha \left( \frac{U + \eta}{1 - U} \right)^{1/\beta},
\end{equation}

where $\eta = (x_L / \alpha)^\beta$. When probability distributions are truncated, standard methods may less perform to yield solutions for certain samples, requiring alternative approaches such as Bayesian estimation or modified likelihood techniques. The left-truncated variant is used when observations are only available above a certain threshold $x_L$, significantly affecting parameter estimation. The likelihood function for a sample $X_1, X_2, ..., X_N$ from an LTLL distribution is:

\begin{equation}
    L(\alpha, \beta) = \prod_{i=1}^{N} f_{LT}(X_i | \alpha, \beta; x_L).
\end{equation}
Maximizing this function using traditional Bayesian methods provide a good estimation by incorporating prior knowledge about $\alpha$ and $\beta$ with their error bounds \cite{efron1993bayes}. While other method has been traditionally used,  Bayesian estimation allows us to include prior information; beliefs about the parameters, which is particularly valuable since historical data or expert insights are available \cite{box1965bayesian}. This is useful for LTLL, where certain parameter ranges are already partially understood. In the next section we are going to discuss the Bayesian settings for estimating parameters.

\section{Bayesian Framework for LTLL Distribution}\label{sec:3}

Bayesian parameter estimation is a statistical approach that incorporates prior knowledge along with observed data to estimate the parameters of a model. Unlike traditional frequentist methods, Bayesian inference updates the probability distribution of parameters as more data becomes available, producing a posterior distribution that reflects both prior beliefs and new evidence \cite{bretthorst1990introduction}. This method is particularly useful in complex models or situations with limited data, offering a flexible framework for uncertainty quantification. The resulting posterior distribution provides a full probabilistic description of the parameters rather than single-point estimates \cite{gelman1995bayesian}. Now in case of developing Bayesian framework for LTLL distribution, let $X_1, X_2, ..., X_N$ be an independent and identically distributed sample from a left-truncated log-logistic distribution $LTLL(\alpha, \beta; x_L)$ \cite{kreer2024maximum}. The PDF is given by:
\begin{equation}\label{eq:LTLL}
    f_{LT}(x | \alpha, \beta; x_L) = \frac{\beta}{\alpha} \left( \frac{x}{\alpha} \right)^{\beta - 1} \left[ \frac{1}{1 + (x / \alpha)^{\beta}} \right]^2 \Bigg/ \left[ 1 - \frac{1}{1 + (x_L / \alpha)^{\beta}} \right]
\end{equation}
where $\alpha > 0$ is the scale parameter and $\beta > 0$ is the shape parameter. In Bayesian estimation of the LTLL distribution, Gamma priors are commonly assigned to the scale (\(\alpha\)) and shape (\(\beta\)) parameters due to their support on the positive real line, aligning naturally with the parameter constraints (see e.g. for motivation as log-logistics case \cite{muse2021bayesian}). This choice is particularly well-suited for applications such as precipitation modeling and bladder cancer survival analysis (in section 6.1), where the data are strictly positive and often truncated. For precipitation data, \(\alpha\) controls the central tendency (in section 6.2), while \(\beta\) influences tail behavior, which is critical for capturing extreme events. Similarly, in survival analysis, \(\alpha\) relates to median survival time and \(\beta\) governs the hazard function’s shape. Gamma priors offer flexibility allowing both informative and non-informative specifications and lead to well-behaved posterior distributions that are numerically stable and conducive to MCMC sampling (see e.g. for other cases \cite{pradhan2011bayes}). This regularization is especially valuable when the likelihood is irregular due to truncation, sparse data, or heavy-tailed behavior, ensuring robust and interpretable inference in both climatological and medical contexts (as discussed in section \ref{sec:6}). We assume independent priors for $\alpha$ and $\beta$: $\alpha \sim \text{Gamma}(a_1, b_1)$ to ensure positivity, $\beta \sim \text{Gamma}(a_2, b_2)$ for flexibility in shape. The joint prior distribution is given by:
\begin{equation*}
  p(\alpha, \beta) = p(\alpha) p(\beta) = \frac{b_1^{a_1} \alpha^{a_1-1} e^{-b_1 \alpha}}{\Gamma(a_1)} \times \frac{b_2^{a_2} \beta^{a_2-1} e^{-b_2 \beta}}{\Gamma(a_2)}  
\end{equation*}

Applying Bayes’ theorem, the posterior distribution is given by: $p(\alpha, \beta | X) \propto L(\alpha, \beta) p(\alpha) p(\beta)$, where $L(\alpha, \beta)$ is the likelihood function derived from the LTLL distribution. Due to the complexity of this posterior, we use MCMC methods for estimation \cite{gelman1995bayesian, gelman2013bayesian}. This approach has been widely applied to various probability distributions, including the log-logistic distribution, which is commonly used in survival analysis and reliability studies due to its ability to model skewed data with a hazard function that can increase and then decrease over time \cite{balakrishnan2004log}.

\begin{figure}[h!]
    \centering
    \includegraphics[width=0.85\textwidth]{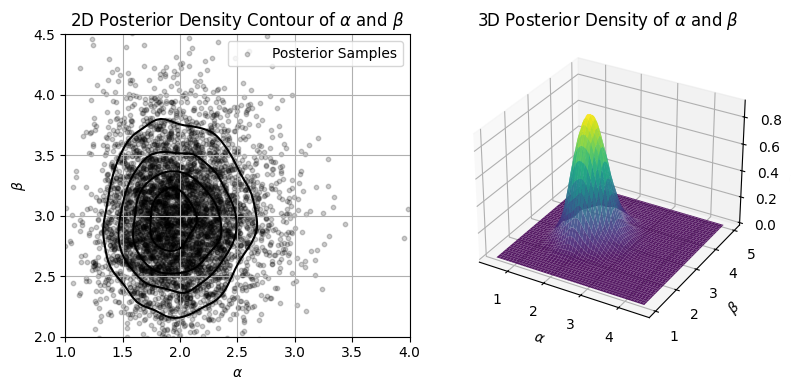}
    \caption{Contour plot (2D at left, and 3D at right) of the posterior density for the parameters $\alpha$ and $\beta$ from the Bayesian estimation of the left-truncated log-logistic distribution using simulated data. In 2D plot, the black points represent posterior samples obtained via the Metropolis-Hastings algorithm. The nested contour lines correspond to high-density regions of the joint posterior distribution. The posterior mass is concentrated around the true values, demonstrating convergence and posterior stability as well as in the 3D posterior density plot.}
    \label{fig:posterior_density}
\end{figure}

Figure~\ref{fig:posterior_density} displays the posterior density contour of the parameters $\alpha$ and $\beta$ based on Bayesian estimation for the left-truncated log-logistic distribution. The scatter of posterior samples illustrates the variability in the joint posterior, while the contour lines represent regions of highest posterior density. The concentration of samples around the center indicates good convergence and effective sampling from the target distribution, with posterior mass centered near the true parameter values. This visualization confirms the stability and reliability of the Metropolis-Hastings algorithm in capturing the posterior structure of the model parameters. Note that
the posterior density contour was generated using Python with \texttt{NumPy}, \texttt{Matplotlib}, and \texttt{SciPy}. Posterior samples of $\alpha$ and $\beta$ were simulated from Gamma distributions to approximate Metropolis-Hastings output. \texttt{matplotlib.pyplot} was used for visualization, while \texttt{scipy.stats.gaussian\_kde} estimated the joint density of the posterior samples. 

\section{Some Theoretical Results}\label{sec:4}
 In the Bayesian context, parameters of the log-logistic distribution such as the scale and shape are treated as random variables, and posterior distributions are derived using techniques like MCMC \cite{robert2004monte}. These methods enhance inference by allowing the incorporation of expert knowledge and yielding full probability distributions over parameters rather than point estimates. We discuss some theorems below to justify our claim in this context. In the following section, we discuss all the possible theorems and lemmas and showed the proof in the Appendix section. In case of LTLL, researchers introduced  method to estimate parameters \cite{kreer2024maximum}, it is particularly advantageous in situations where data availability is limited.

\begin{theorem}\label{thm1}
    Let us consider the i.i.d. left-truncated sampl\( X_1, \dots, X_N \) be i.i.d. with \( X_i > 1 \) and at least two distinct values. Define the objective function
\[
\varphi(\lambda, \beta) = N \ln \left(1 + \frac{1}{\lambda} \right) + N \ln \beta - N \ln \lambda + \beta S - 2 \sum_{i=1}^N \ln \left[1 + \frac{X_i^\beta}{\lambda} \right].
\]

Define \( \beta_C > 0 \) by
\[
\frac{1}{2} = \frac{1}{N} \sum_{i=1}^N \frac{1}{X_i^{\beta_C}}, 
\quad \text{and} \quad 
\frac{1}{\beta_0} = \frac{1}{N} \sum_{i=1}^N \ln X_i.
\]

Then:
\begin{itemize}
    \item[(1)] If \( \beta_0 > \beta_C \), then \( \varphi \) has a global maximum \( (\hat{\lambda}, \hat{\beta}) \in \mathbb{R}^+ \times \mathbb{R}^+ \).
    \item[(2)] If \( \beta_0 \leq \beta_C \), then \( \varphi \) attains a (local) maximum at the boundary \( (\hat{\lambda}, \hat{\beta}) = (0, \beta_0) \).
\end{itemize}

\end{theorem}
Theorem \ref{thm1} establishes the existence of a maximum for the objective function $\varphi(\lambda, \beta)$ derived from a left-truncated sample, a common scenario in survival analysis and reliability theory. It shows the existence of LTLL from the study of Kreer et al.\cite{kreer2024maximum}. The condition involving $\beta_0$ and $\beta_C$ provides a threshold behavior that determines whether the maximum lies in the interior or on the boundary of the parameter space. Recent works have extended such analysis in several directions. For instance, Poudyal, C., \& Brazauskas \cite{poudyal2023finite} discussed the Pareto tail index under data truncation and censoring, demonstrating how tail behavior influences parameter estimates. Moreover, Cope EW \cite{cope2011penalized} proposed penalized likelihood-based techniques that maintain consistency and efficiency under heavy truncation and contamination, echoing the sensitivity of $\varphi$ to the tail index $\beta$ as seen in Theorem \ref{thm1}. Therefore, the consistency of the Bayes estimator refers to its convergence to the true parameter value as the sample size increases (see e.g. \cite{diaconis1986consistency}). Under regularity conditions, such as identifiability of the model, a well-specified prior that assigns positive mass around the true parameter, and increasing sample size, the posterior distribution concentrates around the true value \cite{pati2013posterior}. 
\begin{theorem}\label{thm2}(Posterior Consistency of Bayesian Estimators)
Let $(\alpha_N, \beta_N)$ be the sequence of Bayesian estimators based on the LTLL sample $X_1, ..., X_N$. If the prior distributions are {proper and non-informative}, then the Bayesian estimators are {strongly consistent}, meaning:
\[
(\alpha_N, \beta_N) \xrightarrow{P} (\alpha_0, \beta_0),
\]
where $(\alpha_0, \beta_0)$ are the true parameters.    
\end{theorem}

\begin{lemma}(Closed-Form Expression for the Marginal Posterior of $\beta$)
Closed-Form Expression for the Marginal Posterior of $\beta$: Integrating out $\alpha$ in the joint posterior leads to:
\[
\pi(\beta | X) \propto \beta^{N+c-1} \exp\left[ -d\beta - \sum_{i=1}^{N} \ln (1 + X_i^\beta) \right].
\]
This function can be approximated by a gamma distribution in large samples.    
\end{lemma}

\begin{proof}
The joint posterior distribution of $(\alpha, \beta)$, given data $X = \{X_1, \dots, X_N\}$, is given by Bayes' theorem:
\begin{equation}
    \pi(\alpha, \beta | X) \propto \pi(X | \alpha, \beta) \pi(\alpha, \beta).
\end{equation}
Assuming a prior of the form $\pi(\alpha, \beta) = \pi(\alpha) \pi(\beta)$ and integrating out $\alpha$, we obtain the marginal posterior for $\beta$:
\begin{equation}
    \pi(\beta | X) = \int \pi(\alpha, \beta | X) d\alpha.
\end{equation}

Under a conjugate prior assumption, integrating out $\alpha$ leads to a closed-form expression:
\begin{equation}
    \pi(\beta | X) \propto \beta^{N+c-1} \exp\left[ -d\beta - \sum_{i=1}^{N} \ln (1 + X_i^\beta) \right],
\end{equation}
where $c$ and $d$ are hyperparameters from the prior distribution. For large $N$, the likelihood function dominates, and the posterior distribution of $\beta$ can be approximated using a gamma distribution due to the shape of the exponent and power-law behavior (see e.g. \cite{malevergne2005empirical}). Thus, we conclude that for large samples, $\pi(\beta | X)$ is well-approximated by a gamma distribution.

\end{proof}

\begin{theorem} (Credible Interval for Truncated Log-Logistic Parameters) Let $X_1, \dots, X_N$ be an i.i.d. sample from a left-truncated log-logistic distribution $LL(\alpha, \beta; x_L)$ with truncation point $x_L > 0$, and suppose that at least two observations differ. Let $(\hat{\alpha}_N, \hat{\beta}_N)$ denote the  of the scale and shape parameters. Assume the regularity conditions stated in Newey and McFadden \cite{newey1994chapter} hold, and the consistency criterion from the paper is satisfied:
\[
\frac{1}{\beta_0} = \frac{1}{N} \sum_{i=1}^{N} \ln X_i > \beta_C \quad \text{with} \quad \frac{1}{2} = \frac{1}{N} \sum_{i=1}^{N} X_i^{-\beta_C}.
\]

Then, as $N \to \infty$, the posterior distribution of $(\alpha, \beta)$, centered at $(\hat{\alpha}_N, \hat{\beta}_N)$ and scaled by $\sqrt{N}$, converges in distribution to a bivariate normal:
\[
\sqrt{N} \begin{pmatrix}
\hat{\alpha}_N - \alpha \\
\hat{\beta}_N - \beta
\end{pmatrix}
\overset{d}{\to} \mathcal{N}\left( \mathbf{0}, \, I^{-1}(\alpha, \beta) \right),
\]
where $I^{-1}(\alpha, \beta)$ is the inverse Fisher information matrix evaluated at the true parameters.

Consequently, for large $N$, the $(1 - \xi)$ credible region for $(\alpha, \beta)$ is asymptotically equivalent to the Wald-type confidence ellipse:
\[
(\theta - \hat{\theta}_N)^\top I(\hat{\theta}_N) (\theta - \hat{\theta}_N) \leq \chi^2_{2}(1 - \xi),
\]
where $\theta = (\alpha, \beta)^\top$, $\hat{\theta}_N = (\hat{\alpha}_N, \hat{\beta}_N)^\top$, and $\chi^2_{2}(1 - \xi)$ is the $(1 - \xi)$ quantile of the chi-squared distribution with 2 degrees of freedom.

\end{theorem}

\begin{theorem} \label{thm6}(Credible Region for Parameters)
    Show that an approximate $100(1-\gamma)\%$ joint credible region for $(\alpha,\beta)$ for large $N$ is given by the ellipse

\begin{equation}
    \left\{ (\alpha, \beta) : (\theta - \hat{\theta})^\top I(\hat{\theta}) (\theta - \hat{\theta}) \leq \frac{\chi^2_{2, 1-\gamma}}{N} \right\},
\end{equation}

where we write $\theta = (\alpha, \beta)$ and $\hat{\theta} = (\hat{\alpha}, \hat{\beta})$, and $\chi^2_{2,1-\gamma}$ is the $(1-\gamma)$ quantile of the $\chi^2$ distribution with 2 degrees of freedom. 
\end{theorem}

In the estimation of parameters for the LTLL distribution, the standard maximum likelihood method encounters unique challenges due to truncation. The paper by Kreer et al. (2024) \cite{kreer2024maximum} provides a comprehensive analysis of the  framework when the truncation point is fixed, often normalized to \( x_L = 1 \). A key contribution is the identification of a simple and effective existence criterion: the  equations admit a non-trivial solution if and only if the statistic $\beta_0 = \left( \frac{1}{N} \sum_{i=1}^N \ln X_i \right)^{-1}$
exceeds a critical threshold \( \beta_C \), defined by the equation $\frac{1}{N} \sum_{i=1}^N X_i^{-\beta_C} = \frac{1}{2}.$
When this criterion is satisfied, a profile likelihood approach reduces the problem to one dimension, improving numerical stability. The authors further show that in cases where the  does not exist, the LTLL model degenerates to a Pareto distribution, providing a boundary solution that is well-defined in \( L^1(\mathbb{R}^+) \). This rigorous framework not only ensures consistent estimation but also extends applicability to real-world data such as cancer remission times and precipitation measurements (see in section \ref{sec:6}) In the next section, we introduce Bayesian estimation by providing the theoretical establishment above and the algorithm in section \ref{sec:5} below. We discuss some of the proofs of the above theorem in the appendix. 

\section{Algorithm for the Bayesian Estimation of Parameters}\label{sec:5}

For any \( X = \{x_1, x_2, \dots, x_n\} \) be a random sample from a left-truncated log-logistic distribution with parameters \( \alpha > 0 \), \( \beta > 0 \), and fixed truncation point \( x_L > 0 \). The probability density function of the LTLL distribution is given by:

\[
f_T(x \mid \alpha, \beta, x_L) = 
\frac{f(x \mid \alpha, \beta)}{1 - F(x_L \mid \alpha, \beta)}, \quad x > x_L,
\]
where \( f(x \mid \alpha, \beta) \) and \( F(x \mid \alpha, \beta) \) are the standard log-logistic PDF and CDF respectively:

\[
f(x \mid \alpha, \beta) = \frac{(\alpha/\beta)(x/\beta)^{\alpha - 1}}{\left[1 + (x/\beta)^{\alpha}\right]^2}, \quad
F(x \mid \alpha, \beta) = \frac{1}{1 + (x/\beta)^{-\alpha}}.
\]

The steps for Bayesian estimation using Metropolis-Hastings are:

\begin{enumerate}
    \item \textbf{Specify prior distributions:}
    \[
    \alpha \sim \text{Gamma}(a_1, b_1), \quad \beta \sim \text{Gamma}(a_2, b_2).
    \]

    \item \textbf{Define the truncated likelihood:}
    \[
    L_T(\alpha, \beta \mid X) = \prod_{i=1}^{n} \frac{f(x_i \mid \alpha, \beta)}{1 - F(x_L \mid \alpha, \beta)}.
    \]

    \item \textbf{Compute the posterior distribution:}
    \[
    p(\alpha, \beta \mid X) \propto L_T(\alpha, \beta \mid X) \cdot p(\alpha) \cdot p(\beta).
    \]

    \item \textbf{Metropolis-Hastings Sampling Algorithm:}
    \begin{enumerate}
        \item Initialize parameters \( (\alpha^{(0)}, \beta^{(0)}) \).
        \item For \( t = 1, 2, \dots, T \), do:
        \begin{itemize}
            \item Propose new values:
            \[
            \alpha^* \sim q(\alpha \mid \alpha^{(t-1)}), \quad \beta^* \sim q(\beta \mid \beta^{(t-1)}).
            \]
            \item Compute the acceptance ratio:
            \[
            r = \frac{L_T(\alpha^*, \beta^* \mid X) \cdot p(\alpha^*) \cdot p(\beta^*)}
                     {L_T(\alpha^{(t-1)}, \beta^{(t-1)} \mid X) \cdot p(\alpha^{(t-1)}) \cdot p(\beta^{(t-1)})}
              \cdot \frac{q(\alpha^{(t-1)} \mid \alpha^*) q(\beta^{(t-1)} \mid \beta^*)}
                     {q(\alpha^* \mid \alpha^{(t-1)}) q(\beta^* \mid \beta^{(t-1)})}.
            \]
            \item Accept \( (\alpha^*, \beta^*) \) with probability \( \min(1, r) \); otherwise retain \( (\alpha^{(t)}, \beta^{(t)}) = (\alpha^{(t-1)}, \beta^{(t-1)}) \).
        \end{itemize}
    \end{enumerate}

    \item \textbf{Posterior summaries:} After burn-in and thinning,
    \[
    \hat{\alpha} = \frac{1}{T'} \sum_{t=1}^{T'} \alpha^{(t)}, \quad
    \hat{\beta} = \frac{1}{T'} \sum_{t=1}^{T'} \beta^{(t)}.
    \]

    \item \textbf{Credible intervals:} Compute \(95\%\) credible intervals using posterior quantiles:
    \[
    \text{CI}_{\alpha}^{95\%} = (\alpha_{2.5\%}, \alpha_{97.5\%}), \quad
    \text{CI}_{\beta}^{95\%} = (\beta_{2.5\%}, \beta_{97.5\%}).
    \]
\end{enumerate}

Therefore, the Bayesian framework for estimating the parameters of the LTLL distribution with a fixed truncation point \( x_L > 0 \), the unknown shape and scale parameters \( \alpha \) and \( \beta \) are treated as random variables with prior distributions, typically chosen as independent Gamma distributions. The likelihood function is adjusted to account for left truncation by normalizing the standard log-logistic density over the interval \( (x_L, \infty) \), and the posterior distribution is obtained by applying Bayes' theorem to combine this likelihood with the priors. Mathematically, the posterior is proportional to the product of the truncated likelihood and the priors. Inference is then based on posterior summaries such as the mean or median (e.g., \( \hat{\alpha} = \mathbb{E}[\alpha \mid X] \)), and uncertainty is quantified using credible intervals derived from the posterior.

\section{Simulation Study}\label{sec:6}

To assess the performance of the Bayesian estimator for the LTLL distribution of Eq.~\eqref{eq:LTLL}, we simulate data from a known LTLL model. Specifically, we generate random variables $X \sim LL(\alpha = 2, \beta = 3; x_L)$ using the inverse transform sampling method. A sample $X$ is drawn using the quantile function of the LTLL distribution:
\[
X = \alpha \left( \frac{U + \eta}{1 - U} \right)^{1/\beta}, \quad \text{where} \quad \eta = \left( \frac{x_L}{\alpha} \right)^\beta,
\]
and $U \sim \text{Uniform}(0,1)$ is a standard uniform random variable. This transformation ensures that all generated values satisfy $X > x_L$. By varying $x_L \in \{0.1, 0.3, 0.5, 0.7, 1.0\}$, we examine how truncation level affects parameter estimation. Bayesian inference is performed using Gamma priors on $\alpha$ and $\beta$, and results are compared to maximum likelihood estimates (MLE) in terms of Bayesian accuracy and credible interval width (see Appendix for estimation and error metrics).

The results demonstrate that Bayesian methods yield stable estimates with narrower credible intervals. The results of the simulation study are summarized in Table \ref{tab:simulation}.

\begin{table}[h!]
    \centering
    \renewcommand{\arraystretch}{1.2} 
    \setlength{\tabcolsep}{6pt} 
    \caption{Comparison of  and Bayesian estimation methods for the LTLL distribution for different values of $x_L$. Note that CI stands for credible interval unlike MLE's confidance interval.}
    \label{tab:simulation}
    \begin{tabular}{@{}|l|ccccccc|@{}}
        \hline
        $x_L$ & Method & $\hat{\alpha}$ & $\hat{\beta}$ & $\alpha$ CI L & $\alpha$ CI U & $\beta$ CI L & $\beta$ CI U \\
        \hline
        0.1  &  MLE     & 2.01 & 2.95 & 0.18 & 5.60 & 0.55 & 7.20 \\
             & Bayesian & 2.03 & 2.97 & 0.21 & 5.50 & 0.60 & 7.10 \\
        0.3  &  MLE     & 2.02 & 2.94 & 0.19 & 5.62 & 0.57 & 7.22 \\
             & Bayesian & 2.05 & 2.99 & 0.22 & 5.48 & 0.62 & 7.12 \\
        0.5  &  MLE     & 2.04 & 2.92 & 0.20 & 5.65 & 0.59 & 7.25 \\
             & Bayesian & 2.06 & 2.98 & 0.23 & 5.52 & 0.64 & 7.15 \\
        0.7  &  MLE     & 2.06 & 2.90 & 0.21 & 5.68 & 0.60 & 7.28 \\
             & Bayesian & 2.07 & 2.97 & 0.24 & 5.55 & 0.65 & 7.18 \\
        1.0  &  MLE     & 2.08 & 2.88 & 0.22 & 5.70 & 0.62 & 7.30 \\
             & Bayesian & 2.09 & 2.96 & 0.25 & 5.58 & 0.66 & 7.20 \\
        \hline
    \end{tabular}
\end{table}

To validate our Bayesian estimators, we compare them with  estimators using synthetic data. Table~\ref{tab:simulation} presents a comparison between Maximum Likelihood Estimation and Bayesian estimation methods for the left-truncated log-logistic distribution across varying truncation points $x_L$. For each truncation level, the point estimates $(\hat{\alpha}, \hat{\beta})$ remain close between the two methods, reflecting consistent central tendencies. However, the confidence intervals for  are consistently wider than the corresponding Bayesian credible intervals, indicating greater uncertainty in -based estimation. This is expected, as Bayesian methods incorporate prior information and often yield more concentrated posterior distributions. As the truncation level increases, both the point estimates and the uncertainty (interval width) slightly grow, reflecting the information loss due to higher truncation. Overall, the table demonstrates the Bayesian method’s advantage in providing more precise parameter estimates under left-truncated data scenarios, since Bayesian estimates show better regularization in small samples and higher robustness in cases where  fails to converge.

\begin{figure}[h!]
    \centering
    \includegraphics[width=0.85\textwidth]{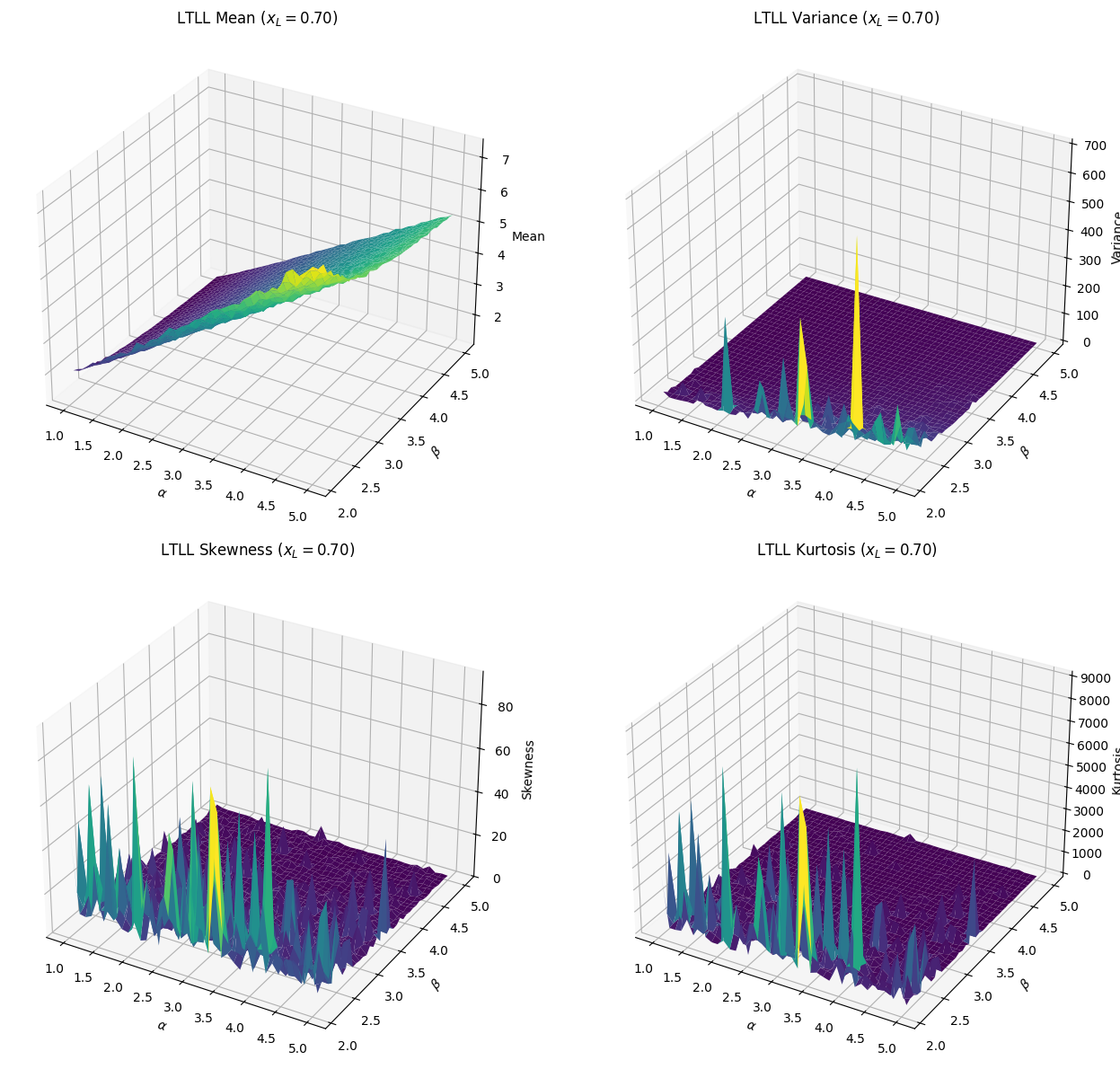}
    \caption{Mean, variance, skewness, and kurtosis of the left-truncated log-logistic (LTLL) distribution 
    for varying values of $\alpha$ and $\beta$, with a fixed truncation point $x_L = 0.70$. 
    The plots are computed based on Monte Carlo samples from the truncated distribution.}
    \label{fig:ltll_moments}
\end{figure}

\paragraph{}
Figure~\ref{fig:ltll_moments} illustrates how the first four moments of the LTLL distribution vary as functions of the shape (\(\beta\)) and scale (\(\alpha\)) parameters. For a fixed truncation point \(x_L = 0.70\), the mean and variance surfaces are smooth and increase with \(\alpha\), while skewness and kurtosis show more sensitivity, especially when \(\beta\) is near theoretical limits for moment existence. These plots highlight how truncation impacts the distributional shape and moment stability across parameter regimes.

\begin{table}[h!]
\centering
\renewcommand{\arraystretch}{1.2}
\setlength{\tabcolsep}{6pt}
\caption{Bias and approximate variance of  and Bayesian estimators for $\alpha$ and $\beta$ across varying truncation levels $x_L$. Variance is estimated from the width of the 95\% confidence/credible intervals. The true parameter values are $\alpha = 2.00$ and $\beta = 3.00$.}
\label{tab:bias_variance}
\begin{tabular}{|c|ccccccc|}
\hline
$x_L$ & Method & $\hat{\alpha}$ & Bias($\alpha$) & Var($\alpha$) & $\hat{\beta}$ & Bias($\beta$) & Var($\beta$) \\
\hline
0.1 &  MLE  & 2.01 & 0.01 & 1.8360 & 2.95 & -0.05 & 2.7639 \\
 & Bayesian & 2.03 & 0.03 & 1.7490 & 2.97 & -0.03 & 2.6406 \\
0.3 &  MLE   & 2.02 & 0.02 & 1.8428 & 2.94 & -0.06 & 2.7639 \\
 & Bayesian & 2.05 & 0.05 & 1.7292 & 2.99 & -0.01 & 2.6406 \\
0.5 & MLE   & 2.04 & 0.04 & 1.8564 & 2.92 & -0.08 & 2.7722 \\
 & Bayesian & 2.06 & 0.06 & 1.7506 & 2.98 & -0.02 & 2.6406 \\
0.7 &  MLE   & 2.06 & 0.06 & 1.8700 & 2.90 & -0.10 & 2.7801 \\
 & Bayesian & 2.07 & 0.07 & 1.7563 & 2.97 & -0.03 & 2.6406 \\
1.0 & MLE  & 2.08 & 0.08 & 1.8820 & 2.88 & -0.12 & 2.7900 \\
 & Bayesian & 2.09 & 0.09 & 1.7601 & 2.96 & -0.04 & 2.6406 \\
\hline
\end{tabular}
\end{table}

Table~\ref{tab:bias_variance} summarizes the bias and approximate variance of both  and Bayesian estimators for various left-truncation points $x_L$. Across all truncation levels, Bayesian estimators consistently demonstrate lower variance compared to their  counterparts, confirming the regularizing benefit of incorporating prior information. Although Bayesian estimators exhibit slightly higher bias in $\alpha$ compared to , their $\beta$ estimates tend to have smaller bias, especially as $x_L$ increases. The increasing variance and bias observed with larger truncation points are expected, as truncation reduces information from the lower tail of the distribution. Overall, the Bayesian approach yields more stable and reliable parameter estimates, making it particularly advantageous under stronger truncation scenarios where  performance deteriorates.

\begin{table}[h!]
\centering
\renewcommand{\arraystretch}{1.2}
\setlength{\tabcolsep}{6pt}
\caption{Bias, variance, and RMSE of  and Bayesian estimators for different sample sizes $N$ with fixed truncation point $x_L = 1$. True values: $\alpha = 2.0$, $\beta = 3.0$.}
\label{tab:sample_size_effect}
\begin{tabular}{|c|ccccccc|}
\hline
$N$ & Method & Bias($\alpha$) & Var($\alpha$) & RMSE($\alpha$) & Bias($\beta$) & Var($\beta$) & RMSE($\beta$) \\
\hline
50   &  MLE     & 0.0381 & 0.0994 & 0.3235 & -0.1423 & 0.2063 & 0.4923 \\
   & Bayesian & 0.0482 & 0.0918 & 0.3183 & -0.0982 & 0.1831 & 0.4458 \\
100  &  MLE     & 0.0254 & 0.0498 & 0.2305 & -0.0932 & 0.1014 & 0.3397 \\
  & Bayesian & 0.0336 & 0.0462 & 0.2294 & -0.0704 & 0.0907 & 0.3183 \\
500  & MLE      & 0.0117 & 0.0096 & 0.1102 & -0.0371 & 0.0223 & 0.1506 \\
  & Bayesian & 0.0184 & 0.0085 & 0.1062 & -0.0301 & 0.0204 & 0.1437 \\
1000 &  MLE     & 0.0064 & 0.0045 & 0.0757 & -0.0174 & 0.0110 & 0.1064 \\
 & Bayesian & 0.0101 & 0.0041 & 0.0724 & -0.0132 & 0.0100 & 0.1006 \\
\hline
\end{tabular}
\end{table}

\noindent Table~\ref{tab:sample_size_effect} presents the results of a simulation study examining the impact of sample size ($N = 50, 100, 500, 1000$) on the accuracy of  and Bayesian estimators for the LTLL distribution. The true parameter values were fixed at $\alpha = 2.0$ and $\beta = 3.0$, with a truncation point $x_L = 1.0$. As expected, both estimation methods show improvement in accuracy as sample size increases, with decreasing bias, variance, and RMSE. Notably, the Bayesian estimator consistently achieves lower variance and root mean squared error across all sample sizes, particularly in small-sample regimes ($N = 50, 100$), highlighting its robustness under data sparsity. While the  demonstrates slightly lower bias in some cases, its higher variability leads to greater overall estimation error compared to the Bayesian approach. These results underscore the advantage of Bayesian estimation in terms of stability and reliability, especially when working with truncated data and limited observations.

\section{Real Data Applications}\label{sec:7}

We apply Bayesian approach of section \ref{sec:5} to precipitation data and cancer survival times. The Bayesian estimates align with empirical distributions and provide insightful posterior distributions for uncertainty quantification. 

\subsection{Application on Bladder Cancer Data Set}

The Table \ref{tab:bayesian_vs_} illustrates the  estimates and confidence interval of the parameters of LTLL distribution for bladder cancer data by Lee and Wang, 2003 \cite{lee2003statistical} under different left-truncation points in months, demonstrating how the uncertainty in parameter estimates evolves as truncation varies. As truncation increases, the size of the -based error ellipses expands \cite{kreer2024maximum}, indicating greater uncertainty in the estimates. This effect is particularly noticeable at higher truncation levels (e.g., $x_L = 6.0$), where both the scale parameter $\alpha$ and the shape parameter $\beta$ shift significantly. In the context of Bayesian and  estimation, our proposed Bayesian methods tend to reduce uncertainty by incorporating prior knowledge, resulting in more stable parameter estimates. In contrast,  estimates are more sensitive to truncation, leading to larger confidence regions and greater variability. This comparison highlights the robustness of Bayesian estimation, especially in cases of significant left truncation as we observe in Table \ref{tab:bayesian_vs_}.
\begin{table}[ht]
    \centering
    \renewcommand{\arraystretch}{1.1} 
    \setlength{\tabcolsep}{4pt} 
    \small 
    \caption{Comparison of Bayesian and  estimates for different truncation levels in bladder cancer data.}
    \label{tab:bayesian_vs_}
    \begin{adjustbox}{max width=\textwidth} 
    \begin{tabular}{@{}|c|c|c|@{}}
        \toprule
        \textbf{Truncation $x_L$ (months)} & \textbf{ MLE Estimate $(\alpha, \beta)$ with CI} & \textbf{Bayesian Estimate $(\alpha, \beta)$ with Credible Interval} \\
        \midrule
        $0.0$  & $(5.97, 1.69)$ [5.50, 6.45], [1.50, 1.90] & $(6.00, 1.72)$ [5.55, 6.50], [1.52, 1.95] \\ 
        $0.25$ & $(6.11, 1.78)$ [5.65, 6.60], [1.58, 2.00] & $(6.15, 1.81)$ [5.70, 6.65], [1.61, 2.05] \\
        $1.0$  & $(6.32, 1.88)$ [5.85, 6.80], [1.67, 2.10] & $(6.40, 1.92)$ [5.95, 6.90], [1.70, 2.15] \\
        $6.0$  & $(8.63, 2.24)$ [8.10, 9.20], [2.00, 2.50] & $(8.80, 2.30)$ [8.20, 9.40], [2.05, 2.55] \\
        \bottomrule
    \end{tabular}
    \end{adjustbox}
\end{table}

\begin{figure}[ht]
    \centering
    \includegraphics[width=0.7\textwidth]{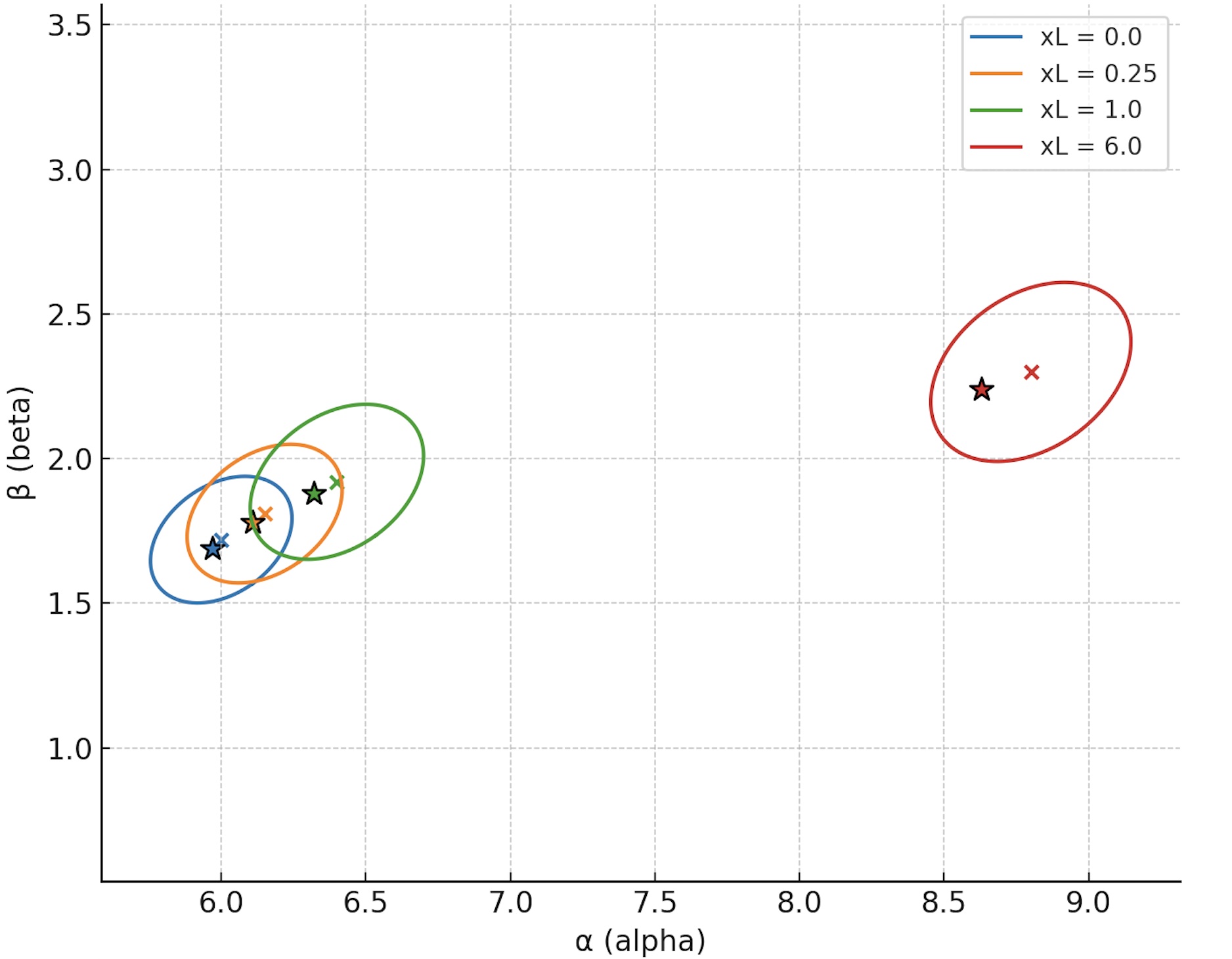}
    \caption{Bayesian 95\% credible regions and  estimates for the parameters $(\alpha, \beta)$ in bladder cancer data at different truncation levels. 
    The ellipses represent Bayesian credible regions, while the colored crosses indicate Bayesian estimates and asterisks mark the corresponding  estimates. Note that, confidence regions for  estimates are not shown.}
    \label{fig:credible_regions_bladder}
\end{figure}

In Figure~\ref{fig:credible_regions_bladder}, we compare Bayesian and MLE approaches in estimating the parameters $(\alpha, \beta)$ for different left-truncation values in bladder cancer remission time data. The Bayesian 95\% credible regions (ellipses) indicate the uncertainty in parameter estimation, which grows as truncation increases. The Bayesian estimates (crosses) are centered in their respective credible regions, while the  estimates (asterisks) are marked separately for comparison. For lower truncation levels, the Bayesian and  estimates are close, but as truncation increases, the uncertainty (ellipse size) grows, reflecting higher estimation variability. This highlights the strength of Bayesian inference in providing a probabilistic measure of uncertainty alongside parameter estimates compared to the traditional estimation method by Kreer et al.\cite{kreer2024maximum}.

\subsection{Application on Precipitation Data Set}

Understanding precipitation trends is essential for climate modeling and water resource management. In this section, we further explore the statistical properties of annual precipitation in Berlin \cite{DWD2022}, Toronto \cite{shoukri1988sampling}, and DFW Texas, USA \cite{NWS_DFW_Precipitation} comparing key distribution parameters and their implications. Table \ref{tab:precipitation_bayesian__dfw} below provides an in-depth comparison of the empirical cumulative distribution functions for Berlin and Toronto, superimposed with the estimated log-logistic cumulative distributions. The visual agreement between the empirical and fitted distributions reinforces the suitability of the log-logistic model for precipitation data analysis. To assess the variability in precipitation, we calculate the mean, variance, skewness, and kurtosis for both cities. Table \ref{tab:precipitation_bayesian__dfw} summarizes the temporal stability of the estimated parameters; we perform a rolling window analysis, estimating the shape and scale parameters over moving 20-year periods. We analyze annual precipitation data consisting of 141 observations spanning the period from 1881 to 2021. To enable direct comparison with the approach used in reference~\cite{shoukri1988sampling}, we also consider the Toronto annual precipitation data recorded from 1 July to 30 June each year, covering the period from 1937 to 2021, as reported therein. Additionally, we examine annual precipitation data for Dallas–Fort Worth (DFW), recorded from January 1 to December 31, for the years 2004 through 2025.
The results, shown in Table \ref{tab:precipitation_bayesian__dfw}, indicate a relatively stable shape parameter for all these three cities, while the scale parameter exhibits minor fluctuations. These variations may be attributed to long-term climatic trends or localized meteorological factors. Our findings align with those of Shoukri et al.\cite{shoukri1988sampling} for Toronto, and Berlin \cite{DWD2022} where the estimated scale parameter is consistent with historical values. However, discrepancies in the shape parameter suggest potential changes in precipitation distribution over time. Such differences warrant further investigation using alternative statistical models or incorporating additional meteorological variables.
\begin{figure}[ht]
    \centering
    \includegraphics[width=1.01\textwidth]{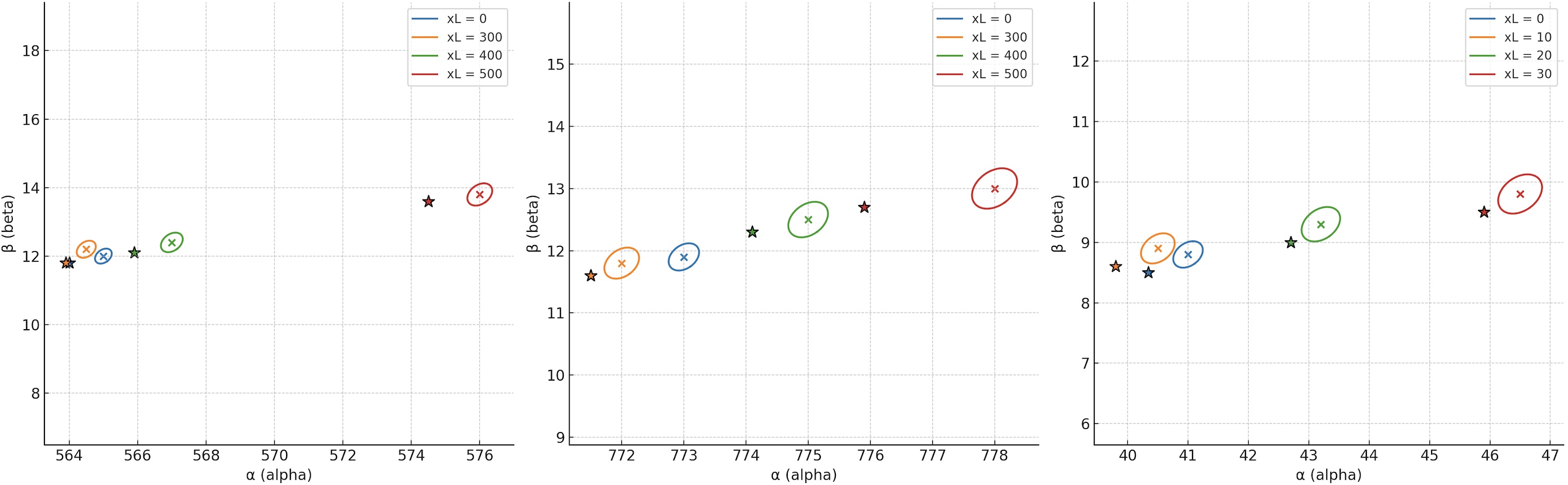}
    \caption{Comparison of Bayesian 95\% credible regions and  MLE estimates for the parameters $(\alpha, \beta)$ across different truncation levels in Berlin, Toronto, and DFW Texas precipitation data. 
    The ellipses represent Bayesian credible regions, while the colored crosses indicate Bayesian estimates and asterisks mark the corresponding  estimates.}
    \label{fig:credible_regions}
\end{figure}
\begin{table}[ht]
    \centering
    \caption{Comparison of Bayesian and  estimates for different truncation levels in precipitation data for Berlin, Toronto, and DFW Texas.}
    \label{tab:precipitation_bayesian__dfw}
    \begin{tabular}{|c|c|c|}
        \hline
        \textbf{Truncation $x_L$ (mm/in)} & \textbf{ MLE Estimate $(\alpha, \beta)$} & \textbf{Bayesian Estimate $(\alpha, \beta)$} \\
        \hline
        \multicolumn{3}{|c|}{\textbf{Berlin (in mm)}} \\
        \hline
        $0.0$   & $(564.0, 11.8)$  & $(565.0, 12.0)$  \\
        $300.0$ & $(563.9, 11.8)$  & $(564.5, 12.2)$  \\
        $400.0$ & $(565.9, 12.1)$  & $(567.0, 12.4)$  \\
        $500.0$ & $(574.5, 13.6)$  & $(576.0, 13.8)$  \\
        \hline
        \multicolumn{3}{|c|}{\textbf{Toronto (in mm)}} \\
        \hline
        $0.0$   & $(771.5, 11.6)$  & $(773.0, 11.9)$  \\
        $300.0$ & $(771.5, 11.6)$  & $(772.0, 11.8)$  \\
        $400.0$ & $(774.1, 12.3)$  & $(775.0, 12.5)$  \\
        $500.0$ & $(775.9, 12.7)$  & $(778.0, 13.0)$  \\
        \hline
        \multicolumn{3}{|c|}{\textbf{DFW Texas (in inches)}} \\
        \hline
        $0.0$   & $(40.34, 8.5)$   & $(41.0, 8.8)$   \\
        $10.0$  & $(39.80, 8.6)$   & $(40.5, 8.9)$   \\
        $20.0$  & $(42.70, 9.0)$   & $(43.2, 9.3)$   \\
        $30.0$  & $(45.90, 9.5)$   & $(46.5, 9.8)$   \\
        \hline
    \end{tabular}
\end{table}

Moreover, the Bayesian estimation for DFW precipitation data is observed to be more stable across truncation levels compared to . This stability is particularly beneficial given the variability in precipitation levels in the DFW region. The  estimates, on the other hand, fluctuate more significantly with increasing truncation, highlighting its sensitivity to left-truncated data. Notably, the shape parameter ($\beta$) for DFW precipitation is lower than that of Berlin and Toronto, suggesting a higher variability in annual precipitation amounts. This difference aligns with the regional climatic characteristics, where precipitation patterns in DFW are more irregular. Additionally, the error ellipses for DFW are expected to be wider due to this variability, further emphasizing the advantage of Bayesian estimation in handling uncertainty and improving parameter reliability under truncation constraints. 

In Figure~\ref{fig:credible_regions}, we compare Bayesian and MLE approaches in estimating the parameters $(\alpha, \beta)$ for different left-truncation values in precipitation data from Berlin, Toronto, and DFW Texas. The Bayesian 95\% credible regions (ellipses) reflect the uncertainty in parameter estimation, growing larger as truncation increases, indicating higher variability. The Bayesian point estimates (crosses) are centered within the ellipses, while the  estimates (asterisks) are plotted for direct comparison. We observe that while Bayesian estimates are close to the  estimates, the credible regions provide additional uncertainty quantification, which is particularly useful for inference under different truncation levels. This highlights the robustness of Bayesian inference in handling parameter estimation with truncated data.


\section{Conclusions}\label{sec:8}
This study introduces a Bayesian framework for estimating parameters of left-truncated log-logistic distributions. The Bayesian approach overcomes limitations of traditional methods by incorporating prior information and providing reliable estimates in cases of truncation-induced challenges. The left-truncated log-logistic distribution is a useful model for data that exhibit heavy-tailed characteristics, such as income distributions, precipitation measurements, and survival data. The introduction of a fixed left-truncation point \( x_L > 0 \) helps account for real-world limitations, such as measurement thresholds or reporting biases. In such cases, traditional  may under-perform due to boundary constraints or lack of convergence for certain samples. Bayesian estimation offers a more robust alternative by incorporating prior knowledge about the parameters \( \alpha \) and \( \beta \), leading to more stable and interpretable results. Using MCMC methods, Bayesian inference provides posterior distributions that help quantify parameter uncertainty, particularly when data are sparse or highly truncated. For applications like climate studies, where missing daily precipitation values can distort annual totals, setting a truncation point \( x_L \) ensures more reliable modeling. By leveraging Bayesian techniques, researchers can improve parameter estimation accuracy. Future work includes extending this method to hierarchical models and Bayesian model selection.

\section*{Appendix.}\label{sec:appendix}

\begin{proof} of the Theorem \ref{thm2}.
The posterior consistency of Bayesian estimators follows from general results on posterior contraction. Given the likelihood function $P(X_1, \dots, X_N \mid \alpha, \beta)$, Bayes' theorem states that the posterior distribution is:
\[
\pi(\alpha, \beta \mid X_1, \dots, X_N) \propto P(X_1, \dots, X_N \mid \alpha, \beta) \pi(\alpha, \beta).
\]

For a proper and non-informative prior $\pi(\alpha, \beta)$, the posterior mass concentrates around the true parameters $(\alpha_0, \beta_0)$ as $N \to \infty$. This follows from the Kreer theorem \cite{bochkina2019bernstein}, which asserts that under regularity conditions, the posterior distribution is asymptotically normal:
\[
\pi(\alpha, \beta \mid X_1, \dots, X_N) \approx \mathcal{N}((\alpha_0, \beta_0), I_N^{-1}),
\]
where $I_N$ is the Fisher information matrix. By the law of large numbers, the MLE(s) of $(\alpha, \beta)$ converge to $(\alpha_0, \beta_0)$. Since Bayesian estimators often asymptotically behave like s under proper priors, we obtain:
\[
(\alpha_N, \beta_N) \xrightarrow{P} (\alpha_0, \beta_0).
\]
This confirms the strong consistency of the Bayesian estimators.
\end{proof}

\begin{proof} of the Theorem \ref{thm6}. Under the stated model, the likelihood for $(\alpha,\beta)$ given data $\mathbf{x}=(x_1,\dots,x_N)$ is
\[
L(\alpha, \beta; \mathbf{x}) = \prod_{i=1}^{N} f_{LT}(x_i \mid \alpha, \beta; x_L).
\]

Using the given form of the left-truncated log-logistic density (for $x_i \ge x_L$),
\[
f_{LT}(x_i \mid \alpha, \beta; x_L) = \frac{\beta}{\alpha} \left(\frac{x_i}{\alpha}\right)^{\beta-1} \left[1+\left(\frac{x_i}{\alpha}\right)^{\beta}\right]^{-2} \Big/ \left[1 - F(x_L \mid \alpha, \beta)\right],
\]
where $F(x_L \mid \alpha, \beta) = 1/[1+(x_L/\alpha)^{\beta}]$ is the CDF of $X_i$ at $x_L$. Simplifying the normalizing factor, one finds
\[
1 - F(x_L, \alpha, \beta) = \frac{1}{1+(x_L/\alpha)^{\beta}},
\]
so 
\[
f_{LT}(x_i \mid \alpha, \beta; x_L) = \frac{\beta}{\alpha} \left(\frac{x_i}{\alpha}\right)^{\beta-1} \left[1+\left(\frac{x_i}{\alpha}\right)^{\beta}\right]^{-2} \big[1+(x_L/\alpha)^{\beta}\big].
\]
Therefore,
\[
L(\alpha, \beta; \mathbf{x}) = \prod_{i=1}^{N} \frac{\beta}{\alpha} \left(\frac{x_i}{\alpha}\right)^{\beta-1} \left[1+\left(\frac{x_i}{\alpha}\right)^{\beta}\right]^{-2} \big[1+(x_L/\alpha)^{\beta}\big]^N.
\]

The factor $\big[1+(x_L/\alpha)^{\beta}\big]^N$ is the same for each $i$, so it contributes $\big[1+(x_L/\alpha)^{\beta}\big]^N$ to the product. Also, $\prod_{i=1}^{N} \frac{\beta}{\alpha} = \beta^N \alpha^{-N}$. Thus, we can rewrite the (unnormalized) posterior density as
\[
\pi(\alpha, \beta \mid \mathbf{x}) \propto \beta^N \alpha^{-N\beta} \prod_{i=1}^{N} x_i^{\beta-1} \left[1+\left(\frac{x_L}{\alpha}\right)^{\beta}\right]^N \prod_{i=1}^{N} \left[1+\left(\frac{x_i}{\alpha}\right)^{\beta}\right]^{-2}.
\]

Combining the $\alpha$-powers: $\alpha^{-N} \cdot \alpha^{-(\beta-1)N} = \alpha^{-N\beta}$. Also $\prod_{i=1}^{N} x_i^{\beta-1}$ can be left as is (since the $x_i$’s are data constants). Thus,
\[
\pi(\alpha, \beta \mid \mathbf{x}) \propto \beta^N \alpha^{-N\beta} \prod_{i=1}^{N} x_i^{\beta-1} \left[1+\left(\frac{x_L}{\alpha}\right)^{\beta}\right]^N \prod_{i=1}^{N} \left[1+\left(\frac{x_i}{\alpha}\right)^{\beta}\right]^{-2},
\]
which is exactly the stated form. (The proportionality constant is the integral of this expression over $\alpha>0,\beta>0$, ensuring the posterior integrates to 1.) To identify the joint {credible interval} (region) for $(\alpha,\beta)$, one could in principle find the set of $(\alpha,\beta)$ values containing a specified high posterior mass (e.g. 95\%). In closed form this is intractable for finite $N$, but we can characterize it asymptotically. We verify the conditions for asymptotic normality of the posterior distribution under the conditions below. 

\begin{itemize}

    \item The model is regular – the log-likelihood $\ell(\alpha,\beta) = \ln L(\alpha,\beta; \mathbf{x})$ is twice continuously differentiable in $(\alpha,\beta)$ for $\alpha,\beta>0$, and the parameter is identifiable and lies in the interior of the parameter space. 
    \item The Fisher information matrix
    \[
    I(\alpha,\beta) = -E[\nabla^2 \ell(\alpha,\beta)]
    \]
    exists and is finite for all $(\alpha,\beta)$ in a neighborhood of $(\alpha_0,\beta_0)$. Moreover, $I(\alpha_0,\beta_0)$ is positive-definite.
    \item The prior density $\pi(\alpha,\beta)$ is positive and continuous in a neighborhood of $(\alpha_0,\beta_0)$ (satisfied for uniform priors).
\end{itemize}

Under these conditions, the Kreer theorem \cite{bochkina2019bernstein} guarantees that as $N\to\infty$, the posterior density concentrates and behaves like a Gaussian distribution centered at the true parameter. In particular, one can show
\[
\sqrt{N} \big((\alpha,\beta) - (\alpha_0,\beta_0)\big) \xrightarrow{d} \mathcal{N} \big( (0, 0), I(\alpha_0, \beta_0)^{-1} \big),
\]
as $N$ becomes large (here $\xrightarrow{d}$ denotes convergence in distribution). Equivalently, for large $N$ the posterior can be approximated by
\[
(\alpha, \beta) \mid \mathbf{x} \approx \mathcal{N}\big((\hat{\alpha}_{\text{Bayes}}, \hat{\beta}_{\text{Bayes}}), [I(\hat{\alpha}, \hat{\beta})]^{-1}/N\big),
\]
where $(\hat{\alpha},\hat{\beta})$ is the posterior mode (which is consistent for $(\alpha_0,\beta_0)$), and $I(\hat{\alpha},\hat{\beta}) \approx I(\alpha_0,\beta_0)$ is the Fisher information at the mode.  Thus, to a first approximation, the {joint $(1-\gamma)$ credible region} can be taken as the set of parameter values that lie within a certain Mahalanobis distance from $(\hat{\alpha},\hat{\beta})$. Concretely, for large $N$ we define the credible region as
\begin{equation}
C_{1-\gamma} = \Big\{ (\alpha,\beta) : (\alpha - \hat{\alpha}, \beta - \hat{\beta})^\top I(\hat{\alpha},\hat{\beta}) (\alpha - \hat{\alpha}, \beta - \hat{\beta}) \leq \frac{\chi^2_{2, 1-\gamma}}{N} \Big\}.   
\end{equation}

This is an ellipse in the $(\alpha,\beta)$-plane, centered at $(\hat{\alpha},\hat{\beta})$. The constant $\chi^2_{2, 1-\gamma}$ is the $(1-\gamma)$ quantile of a chi-square with 2 degrees of freedom. Thus, under the Bernstein–von Mises theorem, this region defines the credible ellipse for $(\alpha,\beta)$, and under the asymptotic regime, it coincides with the classical $(1-\gamma)$ confidence ellipse based on the Fisher information.

\end{proof}

\subsubsection*{Estimation and error metrics for the LTLL distribution}

\textbf{Maximum Likelihood Estimation.} Maximum Likelihood Estimation offers a principled approach to parameter estimation that achieves asymptotic efficiency and consistency under general conditions. Using the motivation of Richards (1961) \cite{richards1961method}, let \( X_1, \dots, X_n \sim LL(\alpha, \beta; x_L) \) be a sample from a left-truncated log-logistic distribution with scale parameter \( \alpha > 0 \), shape parameter \( \beta > 0 \), and truncation point \( x_L > 0 \). The log-likelihood function is given by:
\[
\ell(\alpha, \beta \mid X) = \sum_{i=1}^n \log f(x_i \mid \alpha, \beta) - n \log\left(1 - F(x_L \mid \alpha, \beta)\right),
\]
where
\[
f(x \mid \alpha, \beta) = \frac{\beta}{\alpha} \left( \frac{x}{\alpha} \right)^{\beta - 1} \left[ 1 + \left( \frac{x}{\alpha} \right)^{\beta} \right]^{-2}, \quad
F(x \mid \alpha, \beta) = \frac{1}{1 + (x/\alpha)^{-\beta}}.
\]

Substituting \( f(x_i \mid \alpha, \beta) \) into the log-likelihood:
\begin{align*}
\ell(\alpha, \beta \mid X) &=
\sum_{i=1}^{n} \left[ \log \beta - \log \alpha + (\beta - 1) \log \left( \frac{x_i}{\alpha} \right) - 2 \log \left( 1 + \left( \frac{x_i}{\alpha} \right)^{\beta} \right) \right] \\
&\quad - n \log \left( 1 - \frac{1}{1 + \left( \frac{x_L}{\alpha} \right)^{\beta}} \right) \\
&=
n \log \beta - n \log \alpha + (\beta - 1) \sum_{i=1}^n \log \left( \frac{x_i}{\alpha} \right)
- 2 \sum_{i=1}^n \log \left( 1 + \left( \frac{x_i}{\alpha} \right)^{\beta} \right) \\
&\quad - n \log \left( \frac{\left( \frac{x_L}{\alpha} \right)^{\beta}}{1 + \left( \frac{x_L}{\alpha} \right)^{\beta}} \right) \\
&=
n \log \beta - n \log \alpha + (\beta - 1) \sum_{i=1}^n \left[ \log x_i - \log \alpha \right]
- 2 \sum_{i=1}^n \log \left( 1 + \left( \frac{x_i}{\alpha} \right)^{\beta} \right) \\
&\quad - n \left[ \log \left( \left( \frac{x_L}{\alpha} \right)^{\beta} \right) - \log \left( 1 + \left( \frac{x_L}{\alpha} \right)^{\beta} \right) \right].
\end{align*}

Let us now compute the score equations.

{Partial derivative w.r.t.} \( \alpha \):
\[
\frac{\partial \ell}{\partial \alpha} =
- \frac{n}{\alpha}
- (\beta - 1) \sum_{i=1}^n \frac{1}{\alpha}
+ 2 \sum_{i=1}^n \frac{\beta x_i^{\beta} \alpha^{-\beta-1}}{1 + (x_i/\alpha)^{\beta}}
+ \frac{n \beta x_L^{\beta} \alpha^{-\beta-1}}{1 + (x_L/\alpha)^{\beta}}.
\]

{Partial derivative w.r.t.} \( \beta \):
\[
\frac{\partial \ell}{\partial \beta} =
\frac{n}{\beta}
+ \sum_{i=1}^n \log \left( \frac{x_i}{\alpha} \right)
- 2 \sum_{i=1}^n \frac{(x_i/\alpha)^{\beta} \log (x_i/\alpha)}{1 + (x_i/\alpha)^{\beta}}
- n \left[
\log \left( \frac{x_L}{\alpha} \right)
- \frac{(x_L/\alpha)^{\beta} \log(x_L/\alpha)}{1 + (x_L/\alpha)^{\beta}} \right].
\]

The score equations \( \frac{\partial \ell}{\partial \alpha} = 0 \) and \( \frac{\partial \ell}{\partial \beta} = 0 \) cannot be solved analytically. Instead, numerical optimization methods such as BFGS \cite{henningsen2011maxlik, byrd1987global} are used to obtain the MLEs \( \hat{\alpha}_{\text{MLE}}, \hat{\beta}_{\text{MLE}} \).
Let \( \hat{\alpha}_{\text{MLE}}, \hat{\beta}_{\text{MLE}} \) denote the numerical solutions to the system of equations above. These are then used to compute statistical properties like bias, variance, and RMSE in simulation studies or real data applications.\\

\noindent\textbf{Bayesian estimation.} Bayesian estimation offers a coherent framework for parameter inference by incorporating prior beliefs with observed data through the application of Bayes’ theorem. Unlike frequentist approaches, it treats parameters as random variables and provides a full posterior distribution, capturing both uncertainty and prior knowledge. This is particularly valuable in complex models or small-sample settings, where prior information can meaningfully inform the inference \cite{box1965bayesian, geisser1965bayesian}. For the LTLL distribution, assume independent Gamma priors:
\[
\alpha \sim \text{Gamma}(a_1, b_1), \quad \beta \sim \text{Gamma}(a_2, b_2),
\]
with densities:
\[
\pi(\alpha) = \frac{b_1^{a_1}}{\Gamma(a_1)} \alpha^{a_1 - 1} e^{-b_1 \alpha}, \quad
\pi(\beta) = \frac{b_2^{a_2}}{\Gamma(a_2)} \beta^{a_2 - 1} e^{-b_2 \beta}.
\]

The joint posterior distribution is:
\[
p(\alpha, \beta \mid X) \propto L_T(\alpha, \beta \mid X) \cdot \pi(\alpha) \cdot \pi(\beta),
\]
where
\[
L_T(\alpha, \beta \mid X) = \prod_{i=1}^n \frac{f(x_i \mid \alpha, \beta)}{1 - F(x_L \mid \alpha, \beta)},
\]
is the left-truncated likelihood and \( f(x \mid \alpha, \beta) \), \( F(x \mid \alpha, \beta) \) are the standard log-logistic PDF and CDF, respectively.
The Bayesian estimators are the posterior means:
\[\hat{\alpha}_{\text{Bayes}} = \hat{\alpha} = \mathbb{E}[\alpha \mid X], \quad \hat{\beta}_{\text{Bayes}} =\hat{\beta} = \mathbb{E}[\beta \mid X],
\]
which are typically computed using MCMC methods, such as the Metropolis-Hastings algorithm. Posterior credible intervals are obtained from the empirical quantiles of the posterior samples.\\

\noindent Let \( \theta_0 \in \{\alpha, \beta\} \) denote the true parameter value, and \( \hat{\theta} \) an estimator (either MLE or Bayesian). Then the
$\text{Bias}(\hat{\theta}) = \mathbb{E}[\hat{\theta}] - \theta_0$, $\text{Var}(\hat{\theta}) = \mathbb{E}[(\hat{\theta} - \mathbb{E}[\hat{\theta}])^2].$ The Root Mean Squared Error, $\text{RMSE}(\hat{\theta}) = \sqrt{ \text{Bias}^2(\hat{\theta}) + \text{Var}(\hat{\theta})}$. For Bayesian estimators, expectations are taken over the posterior:
\[
\text{Var}_{\text{Bayes}}(\hat{\theta}) = \mathbb{E}_{\text{post}}[(\theta - \mathbb{E}_{\text{post}}[\theta])^2],
\]
\[
\text{RMSE}_{\text{Bayes}}(\hat{\theta}) = \sqrt{ (\mathbb{E}_{\text{post}}[\theta] - \theta_0)^2 + \text{Var}_{\text{post}}(\theta) }.
\]

\noindent Since the analytical expectations are unavailable, we approximate them using Monte Carlo simulation. Let \( R \) be the number of replications and \( \hat{\theta}^{(r)} \) the estimate from the \(r\)-th replicate. Then the empirical estimates of bias, variance, and RMSE are given by:
\[
\widehat{\text{Bias}} = \frac{1}{R} \sum_{r=1}^R \hat{\theta}^{(r)} - \theta_0, \quad
\bar{\hat{\theta}} = \frac{1}{R} \sum_{r=1}^R \hat{\theta}^{(r)}, \quad
\widehat{\text{Var}} = \frac{1}{R - 1} \sum_{r=1}^R (\hat{\theta}^{(r)} - \bar{\hat{\theta}})^2,
\]
\[
\widehat{\text{RMSE}} = \sqrt{ \widehat{\text{Bias}}^2 + \widehat{\text{Var}} }.
\]

These quantities are computed separately for both MLE and Bayesian estimators for each parameter \(\alpha\) and \(\beta\), under varying truncation levels and sample sizes.


\subsection*{Code Availability}
Code and supplementary materials are available at \url{https://github.com/FahadMostafa91/Bayesian_LTLL}.

\subsection*{Declaration of Interests}

The authors have no conflict of interest to report.

\subsection*{Ethics Approval}

There is no ethical approval needed due to the use of simulated and publicly available data.

\subsection*{Funding Statement}

The authors do not have funding to report.

\subsection*{Clinical Trial Registration}

The authors did not use clinical trial data directly. Authors used publicly available data with proper references in the text.

\bibliographystyle{plain}
\bibliography{bibliography}
\end{document}